\documentclass[letterpaper, 1 pt, conference]{IEEEtran}
\IEEEoverridecommandlockouts
\usepackage{fancyhdr}
\usepackage[utf8]{inputenc}
\usepackage[T1]{fontenc}
\usepackage{cite}
\makeatletter
\let\NAT@parse\undefined
\makeatother
\usepackage[colorlinks,citecolor=black, linkcolor=black]{hyperref}
\usepackage{graphics}
\usepackage{epsfig}
\usepackage{mathptmx}
\usepackage{amsmath}
\usepackage{amssymb}
\usepackage{makecell}
\usepackage{multirow}
\usepackage{graphicx}
\usepackage{subfigure}

\usepackage{xspace}
\makeatletter
\DeclareRobustCommand\onedot{\futurelet\@let@token\@onedot}
\def\@onedot{\ifx\@let@token.\else.\null\fi\xspace}

\def\eg{\emph{e.g}\onedot} 
\def\ie{\emph{i.e}\onedot}

\def\etal{et al\onedot}
\makeatother

\title{\LARGE \bf DsMCL: Dual-Level Stochastic Multiple Choice Learning for \\ Multi-Modal Trajectory Prediction}

\fancypagestyle{FirstPage}{
\fancyhead[L]{
  \copyright~2020 IEEE. Personal use of this material is permitted. Permission from IEEE must be obtained for all other uses, in any current or future media, including reprinting/republishing this material for advertising or promotional purpose, creating new collective works, for resale or redistribution to servers or lists, or reuse of any copyrighted component of this work.}
\fancyfoot[]{}
}

\author{
    Zehan Wang$^{1}$, Sihong Zhou$^{1}$, Yuyao Huang$^{1}$,  Wei Tian$^{1, *}$
    \thanks{$^{1}$ School of Automotive Studies and Institute of Intelligent Vehicles, Tongji University, Shanghai, China}
    \thanks{$^{*}$ Corresponding author. E-mail: tian\_wei@tongji.edu.cn huangyuyao@tongji.edu.cn}
}

\begin{document}

\maketitle
\thispagestyle{FirstPage}
%\pagestyle{empty}

%%%%%%%%%%%%%%%%%%%%%%%%%%%%%%%%%%%%%%%%%%%%%%%%%%%%%%%%%%%%%%%%%%%%%%%%%%%%%%%%%

\begin{abstract}

For both driving safety and efficiency, automated vehicles should be able 
to predict the behavior of surrounding traffic participants in a dynamic environment. To accomplish such a task, trajectory prediction is the key. Although many researchers have been engaged in this topic, it is still challenging. One of the important and inherent factors is the multi-modality of vehicle motion. At present, related researches have more or less shortcomings for multi-modal trajectory prediction, such as requiring explicit modal labels or multiple forward propagation caused by sampling. In this work, we focus on overcoming these issues by pointing out the dual-levels of multi-modal characteristics in vehicle motion and proposing the dual-level stochastic multiple choice learning method (named as DsMCL, for short). This method does not require modal labels and can implement a comprehensive probabilistic multi-modal trajectory prediction by a single forward propagation. By experiments on the NGSIM and HighD datasets, our method has proven significant improvement on several trajectory prediction frameworks and achieves state-of-the-art performance.

\end{abstract}

%%%%%%%%%%%%%%%%%%%%%%%%%%%%%%%%%%%%%%%%%%%%%%%%%%%%%%%%%%%%%%%%%%%%%%%%%%%%%%%%%

\section{INTRODUCTION}

In recent years, the field of autonomous driving has been booming. In automated vehicles, the prediction module plays a pivotal role as a bridge between the perception and the planning module. Without prediction, even perfect perception or planning would never be safe and efficient in practical scenarios. Nowadays, researches on predicting the behavior of traffic participants have received increasing attention.

%\begin{figure}[tbp]
%    \centering
%    \includegraphics[width=60mm]{Pic/Fig1.png}
%    \caption{In this case, the possible action of the target vehicle (black) may be braking (yellow lines) or overtaking (green lines) in order to avoid collision with the front blue vehicle.} 
%    \label{Fig1} 
%    \end{figure}

Currently, plenty of researches focus the prediction task on the design of complex functions to model interactions~\cite{6}~\cite{12}~\cite{13}. However, the fundamental problem that persistently exists but is not well-solved is the multi-modality of vehicle motion. Because of the disparate driving behaviors under the same condition, the vehicle motion has a strong multi-modal characteristic. At present, studies related to dealing with the multi-modality problem can be divided into two kinds. The first one requires predefined modes. For example, Deo~\etal defined six maneuver classes in highway scenes. For each class they generated one corresponding trajectory of bivariate Gaussian distribution~\cite{7}. In the second kind, instead of predefinition, researchers prefer to learn the multi-modality of vehicle trajectories, \eg, by adding latent variables into the network~\cite{10}. However, both kinds of methods have disadvantages. The first kind requires to label trajectory modes in the dataset~\cite{7}, which is troublesome since human annotation cannot guarantee the optimality and even the correctness of labels. In the second kind, multiple predictions are generated by repeatedly sampling and propagation~\cite{10}. Since the sampling may be incomplete or suffer from mode collapse, it is difficult to ensure prediction results to cover all possible modes.

Regarding these issues, this paper gives an in-depth analysis of the dual-level multi-modal characteristics of vehicle motion and proposes a novel stochastic multiple choice learning approach based on that to improve the prediction performance on multi-modal vehicle trajectories. The contributions of this work are as follows:

\begin{itemize}
    \item We decompose the multi-modality of vehicle trajectory into two levels: the intention and the motion. Such a dual-level decomposition well describes the real driving characteristics and is able to predict minor dynamic changes while maintaining primary modes.
    \item We propose a dual-level stochastic multiple choice learning method (dubbed DsMCL), which does not require human-annotated mode labels and can implement a comprehensive probabilistic multi-modal trajectory prediction in a single forward propagation.
    \item We evaluate our proposed method on both NGSIM and HighD datasets. The experimental results demonstrate that our method yields significant improvement on several trajectory prediction frameworks and achieves state-of-the-art performance.
\end{itemize}

%%%%%%%%%%%%%%%%%%%%%%%%%%%%%%%%%%%%%%%%%%%%%%%%%%%%%%%%%%%%%%%%%%%%%%%%%%%%%%%%%

\section{RELATED WORKS}

Trajectory prediction is to predict the future motion of traffic participants by means of regression. A comprehensive review of this field can be referred to works~\cite{2}~\cite{3}. 

\textbf{Deep Learning Methods}  With the success of many revolutionary Deep Neural Network (DNN) frameworks, the era of Deep Learning has arrived~\cite{4}. In recent years, LSTM encoder-decoder architecture has shown excellent performance in natural language processing~\cite{5}. Since the trajectory prediction can also be viewed as a sequence generation problem, the LSTM encoder-decoder architecture was introduced into the field of trajectory prediction and became one of the most popular prediction frameworks. For instance, Alahi et al. designed Social LSTM to predict the trajectory and firstly proposed the Social Pooling structure to model interactions~\cite{6}. Based on that, Deo et al. proposed the Convolutional Social Pooling to make multi-modal trajectory predictions~\cite{7}. Gupta et al. proposed a novel pooling mechanism to further model interactions in a pair-wise manner and achieved more practical results by employing generative adversarial networks~\cite{8}.

\textbf{Methods for Handling Multi-modality}  
While progresses have been seen in the development of prediction models, challenges still exist. One of the important issues is how to solve the multi-modality. Current methods which consider the multi-modality can be divided into two types according to whether modes should be defined in advance.

The first type requires to define all possible modes of vehicle motion, and generates one corresponding trajectory for each mode~\cite{9}. Deo et al. defined six separate maneuver classes in highway scenarios and designed CS-LSTM network for multi-modal trajectory prediction~\cite{7}. All maneuver classes are fed into the regression network in parallel in a coded form to generate corresponding trajectories. Although this type of method can yield multi-modal results in a single forward propagation, the human annotation cannot guarantee the optimality and even the correctness of labels.

The other type does \textbf{not} need to define modes in advance but attempts to learn the multi-modality directly from data~\cite{8}. Tang et al. designed the MFP-k network, where discrete latent variables were added to an EM algorithm framework to learn the multi-modality~\cite{10}. Although this type of method does not require human-annotated labels, it propagates forward multiple times to generate multi-modal results. Additionally, predictions generated by the sampling procedure could not guarantee to cover all possible modes. 

Our work is inspired by the philosophy of Stochastic Multiple Choice Learning (sMCL), which is adopted for training diverse deep ensembles~\cite{1} and implemented in CNN models for trajectory prediction~\cite{11}. Leveraging this philosophy, we propose a label-free prediction approach, which can well handle the multi-modality of vehicle trajectory. The prediction results of our approach can be generated in a single forward propagation and cover all possible future maneuvers of the target vehicle.

%%%%%%%%%%%%%%%%%%%%%%%%%%%%%%%%%%%%%%%%%%%%%%%%%%%%%%%%%%%%%%%%%%%%%%%%%%%%%%%%%

\begin{figure*}[tbp]
\includegraphics[width=\textwidth]{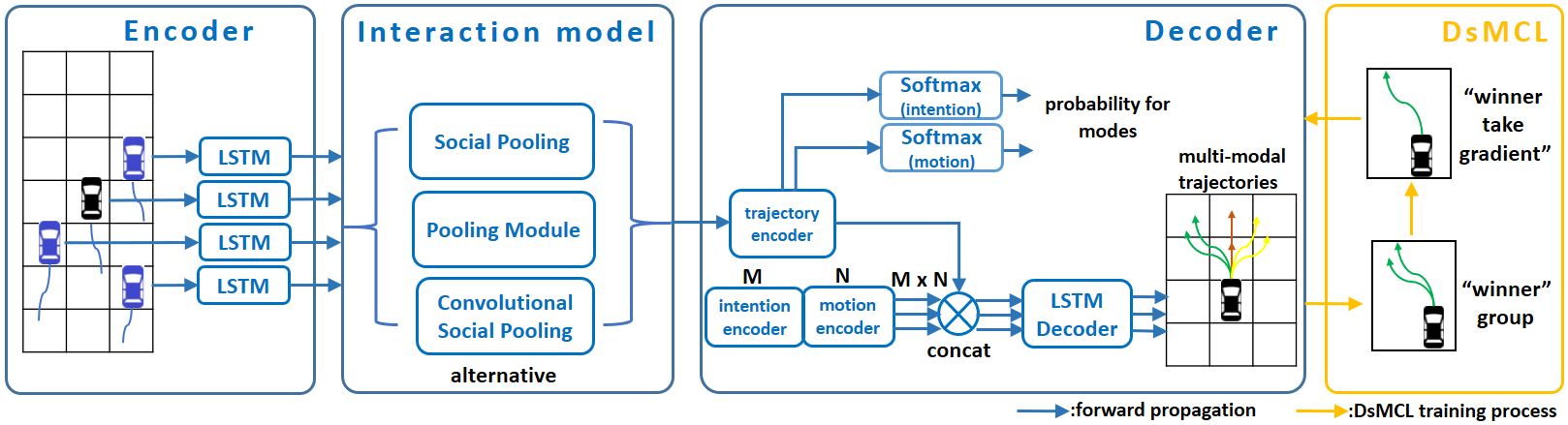}
    \caption{\textbf{DsMCL Multi-modal Trajectory Prediction Framework:} it is composed of Encoder, Interaction model and Decoder. The training process is by the dual-level stochastic multiple choice learning strategy, \ie, "winner-take-gradient".}
    \label{Fig2}
\end{figure*}

\section{DsMCL for Multi-Modal Trajectory Prediction}

In this section, we describe the proposed method for predicting vehicles' multi-modal trajectories. Firstly, we give the problem definition and introduce the dual-level multi-modal characteristics of vehicle motion. Thereafter, we present the details of our DsMCL training method.

\subsection{Problem Definition}

In this paper, we assume that perception systems (\eg LIDARs and cameras) installed on the self-driving car have already accomplished detection and tracking tasks for all surrounding vehicles. The trajectory prediction can be viewed as a sequence generation problem by discretizing the motion of those vehicles with a time step $\Delta t$. Here we denote the position of vehicle $i$ at time $t$ as $X_i^t=(x^t_i, y^t_i)$, where $(x, y)$ are Frenet coordinates with y-axis along the longitudinal direction of the road. Here we make predictions for one vehicle (called as the \textit{target vehicle}) in the surrounding at a time. At each time stamp $t$, the prediction system is given with historical trajectories of the target vehicle $O$ and all its $L$ neighbors as inputs. Therefore, the inputs of the prediction system can be expressed as
\begin{equation}
   \mathbf{X}=\{X_O, X_1, X_2,..., X_L\},
\end{equation}
where $X_i=\left[(x^{t-t_n}_i,y^{t-t_n}_i),(x^{t-t_n+\Delta t}_i,y^{t-t_n+\Delta t}_i),...,(x^{t}_i,y^{t}_i)\right]$ denotes the trajectory of vehicle $i$. The real future trajectory of vehicle $O$ is denoted as
\begin{equation}
   {\bar{Y}_{O}}=\left[(\overline{x}^{t+\Delta t},\overline{y}^{t+\Delta t}),...,(\overline{x}^{t+t_f},\overline{y}^{t+t_f})\right].
\end{equation}
Since prediction systems adopted in this paper can predict multiple feasible future trajectories, we denote the set of predicted trajectories as  $\{Y^j_O\}$, where $j = 1, 2, \cdots, M\times N$. 

\subsection{Dual-level Multi-modal Characteristics of Vehicle Motion}

Moving vehicles have strong multi-modal characteristics. This poses great challenges to the prediction model because it requires the prediction model to perform a one-to-many mapping rather than a one-to-one mapping. To completely describe multi-modalities, we divide the multi-modal characteristics of vehicle motion into two levels. The first level is about the multi-modality of driving \textit{intention} which includes going straight, turning left or right. This level can be considered as a macro perspective. The other is about the multi-modality of trivial, stylized, or accidental \textit{motion} under the same driving intention, such as turning at different speeds or steering angles. This level is regarded as a micro perspective. To construct a network that can generate the described dual-level multi-modal trajectories, we design the specialized training process introduced below, which can be easily integrated into standard trajectory prediction frameworks.

\subsection{DsMCL Multi-modal Trajectory Prediction Framework}

The DsMCL multi-modal trajectory prediction framework is based on the CS-LSTM~\cite{7} and shown in Fig.~\ref{Fig2}. It is a standard trajectory prediction framework consisting of three modules: Encoder, Interaction Model and Decoder. In this framework, track histories of the target and its surrounding vehicles are encoded by LSTMs independently in the Encoder. Then, all the encodings are given to the Interaction Model to deal with motion inter-dependencies between the target vehicle and its neighbors. Finally, the Interaction Model passes both the historical encodings and the interaction information to the Decoder to generate future trajectories along with their probabilities.

Based on the concept of dual-level multi-modal characteristics of vehicle motion, we define $M$ intention modes and $N$ motion modes in this paper. In order to generate trajectories for different modes, we modify the input of LSTM Decoder by concatenating it with two additional one-hot vectors respective to $M$ and $N$ dimensions. We run the forward propagation of the network to obtain $M$ groups of different intention modes and each group contains $N$ trajectories referring to motion modes. Thus, a total of $M\times N$ trajectories can be generated with each corresponding to one combination of intention and motion mode. Although all of the $M\times N$ trajectories can be reasonable, only one of them is selected to calculate the loss with the ground-truth during training. The selected trajectory is referred to as the "winner", as these $M\times N$ trajectories compete with each other in loss calculation and back-propagation. Such a case is called as "winner-take-gradient"~\cite{1}. With this training strategy, the network can learn the multi-modal distribution in the corresponding probability space, where different candidate trajectories represent different mode combinations, according to the competitive arbitration rules.

\subsection{Arbitration of the Dual-Level Mode Selection}

In this section, we introduce how to select candidates for the training procedure in DsMCL. The entire arbitration process is divided into two phases. 

\textbf{Determination of Intention Mode: } The first step is to find out the group ID of the "winner", \ie, to determine the current intention mode. Based on the fact that intention modes are mainly related to vehicles' lateral movement and easy to distinguish over time, we choose the group of trajectories with the smallest lateral deviation from the groundtruth and denote the "winner" intention mode as $m^*$. The deviation is calculated along the $x$-axis between the farthest points of predicted trajectories and the groundtruth. Thus, the selection process can be interpreted as
\begin{equation}
    m^*=\mathop{\arg\min}_{m\in M}\left[\sum_{n\in N}|x^{t+t_f}_{m,n}-\overline{x}^{t+t_f}|\right],
\end{equation}

\textbf{Determination of Motion Mode: } 
Next, we need to determine the motion mode $n^*$ of the "winner" within the intention group $m^*$. Considering that  motion modes under the same intention can differ in the speed or slight lateral deviation, we use the Average Displacement Error (ADE) to distinguish between different motion modes. Here we choose the trajectory with the smallest ADE and denote the "winner" motion mode as $n^*$. The selection process can be interpreted as
\begin{equation}
    L_{ADE}(Y_{m^*,n},\overline{Y})= \frac{1}{H}\sum^{H}_{h=1} \left\|(\overline{x}^{h},\overline{y}^{h})-(x_{m^*,n}^{h},y_{m^*,n}^{h})\right\|,
\end{equation}
\begin{equation}
n^*=\mathop{\arg\min}_{n\in N}L_{ADE}(Y_{m^*,n},\overline{Y}),
\end{equation}
where $H$ denotes the length of prediction sequence.

\subsection{Loss Calculation}
After selecting the winner trajectory to take the gradient, the regression loss for data samples is defined as $L_{ADE}(Y_{m^*,n^*},\overline{Y})$.
%\begin{equation}
%    L_{ADE}(Y_{m^*,n^*},\overline{Y})= %\frac{1}{H}\sum^{H}_{h=1} %\left\|\overline{\gamma}^{t+h\cdot\Delta %t}-\gamma_{m^*,n^*}^{t+h\cdot\Delta t} \right\|,
%\end{equation}
%where $H$ denotes the length of prediction sequence, $\overline{\gamma}^{t+h\cdot\Delta t}$ and $\gamma_{m^*,n^*}^{t+h\cdot\Delta t}$ represent $(\overline{x}^{t+h\cdot\Delta t},\overline{y}^{t+h\cdot\Delta t})$ and $(x_{m^*,n^*}^{t+h\cdot\Delta t},y_{m^*,n^*}^{t+h\cdot\Delta t})$ respectively.

From the above equation, we can see that only the most promising trajectory proposal is viewed as the "correct mode" and will be selected during training. This enables the output of each mode to specialize on a subset of the data, and thus guarantees the success of multi-modal trajectory prediction.

We also train an additional branch for estimating the probability of each predicted trajectory. We regard it as a classification task. This is useful at the inference stage to filter out impractical predictions. Thus, we push the probability of the "winner" as close as possible to 1, and the probabilities of others to 0. During training, we update all probability outputs while the trajectory output is only updated for the "winner". The classification loss $L_{class}$ is in a cross-entropy form and formulated as
\begin{equation}
    L_{class}=-\log P_{m^*,n^*}.
\end{equation}
And the final loss function can be defined as
\begin{equation}
    L_{DsMCL}=L_{class}+\alpha L_{ADE}(Y_{m^*,n^*},\overline{Y}),
\end{equation}
where the hyper-parameter $\alpha$ is a trade-off between two losses.

%%%%%%%%%%%%%%%%%%%%%%%%%%%%%%%%%%%%%%%%%%%%%%%%%%%%%%%%%%%%%%%%%%%%%%%%%%%%%%%%%

\section{EXPERIMENTS}
\subsection{Datasets}

\textbf{NGSIM}~\cite{14}: It is a huge amount of publicly available datasets captured in 2005 and widely used for trajectory prediction evaluation~\cite{7}~\cite{9}~\cite{10}. Same as in~\cite{7}, we also choose the US-101 and I-80 datasets for our experiments. Each of the datasets consists of trajectories of real freeway traffic and is captured at 10 Hz over a period of 45 minutes.   

\textbf{HighD}~\cite{15}: It is a new dataset built in 2017 and 2018. The data was captured from an aerial perspective at 25 Hz and consists of vehicle trajectories on six different German highways. We also use this dataset to further demonstrate the effectiveness of our proposed DsMCL training method.

%\begin{figure}[tbp]
%   \centering
%    %\frame{
%    \setlength{\fboxsep}{10pt}%
%    \setlength{\fboxrule}{0.5pt}%
%    \fbox{
%    \includegraphics[width=55mm]{Pic/Fig3.png}
%    }
%    \caption{Comparison of two prediction networks.}
%    \label{Fig3}
%\end{figure}

\subsection{Evaluation Metrics}

Since trajectory probabilities are available in multi-modal trajectory prediction, a feasible evaluation method is to calculate the Root Mean Square Error (RMSE) between the most probable trajectory and the groundtruth~\cite{7}. However, Tang \etal have pointed out that the RMSE is not a good metric for multi-modal distributions~\cite{10}.In our experiment, we choose minRMSE as our evaluation metric. We filter out trajectories with very low probabilities and then calculate minRMSE among the remaining ones with the groundtruth. In a word, all the predicted trajectories with probabilities higher than a predefined threshold (set to $0.1$ in our experiment) will be considered and compared with the groundtruth. Also, this metric has been proven more appropriate for evaluation of multi-modal trajectory prediction~\cite{10}~\cite{11}.

\subsection{Models and Training Details}

In experiments, we adopt the CS-LSTM~\cite{7} framework as the backbone.
In the original training procedure of CS-LSTM, the mode classifier is learned from rule-based groundtruth, and back-propagation is only taken on the trajectory representing the groundtruth. 
Here, we adopt our DsMCL training method for CS-LSTM and denote the new method as CS-LSTM-DsMCL.
In the original work of CS-LSTM, it defines 6 discrete modes (lateral modes: straight, left and right turn; longitudinal modes: normal, braking). For a fair comparison, in our approach we set parameter $M$ to 3, and $N$ to 2. Thus, the total number of modes is same.
Additionally, we extract 8s of each trajectory sample, with the first 3s for the historical input and the next 5 seconds for the future prediction, which is same as in~\cite{7}. Our model is implemented by Pytorch~\cite{16}. The learning rate is set to 1e-3 during training and we use Adam~\cite{17} for optimization.

\subsection{Qualitative Analysis}

To demonstrate the effectiveness of DsMCL training for multi-modal trajectory prediction, we evaluate the CS-LSTM-DsMCL framework on the NGSIM dataset. For a better understanding of the influence of each level in the multi-modal characteristics, in the first experiment, only the intention level is defined and the number of network outputs is thus set to 3. We also adopt the intention mode arbitration process to determine the "winner" during training.

\begin{figure}[tbp]
    \centering
    \includegraphics[width=75mm]{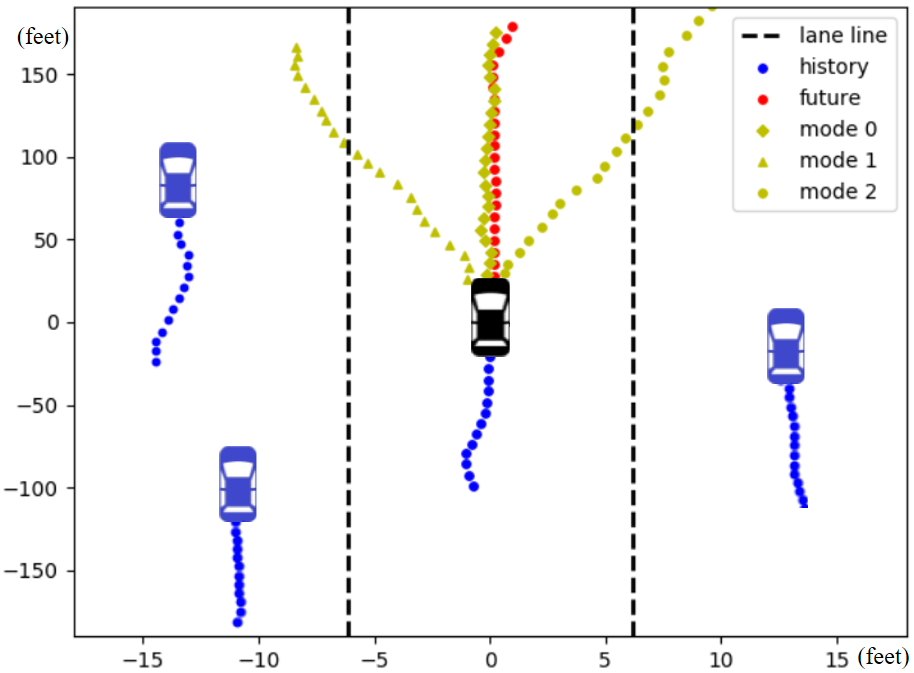}
    \caption{Prediction with intention modes.}
    \label{Fig4}
\end{figure}

Qualitative results are shown in Fig.~\ref{Fig4}.  We can see that there is a large lateral deviation among three predicted trajectories. They can be considered to represent the intention modes of driving straight, turning left or right respectively. Therefore, we can conclude that although we never manually label the modes, the network is still able to learn from the data to cluster trajectory predictions into those most evident modes.

\begin{figure}[tbp]
    \centering
    \includegraphics[width=75mm]{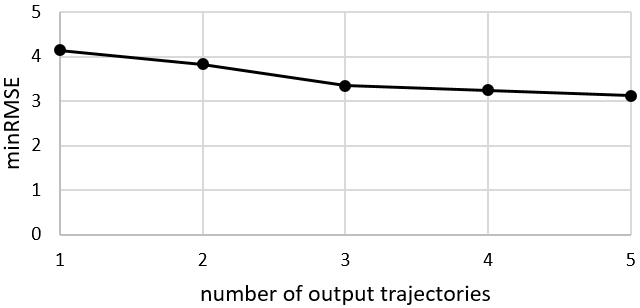}
    \caption{Comparison of prediction performance with different intention mode numbers.}
    \label{Fig5}
\end{figure}

In Fig.~\ref{Fig5}, we show the minRMSE of varied number of output trajectories. It can be seen that the more trajectories we output, the higher the accuracy in terms of minRMSE is. This is because that with the finer division of modes, the output trajectories are more likely to get close to the groundtruth. Moreover, we find out that when the number of intention modes exceeds 3, the accuracy increase slows down. We believe that this fact is related to the used traffic scene data, in which the intention modes on the highway are mainly going straight, turning left and right. 
Thus, we set the optimal output number of the network to 3.

\begin{figure}[tbp]
    \centering
    \subfigure[Under correct intention.]{
        \includegraphics[width=75mm, trim = 0mm 0mm 0mm 0mm, clip=true]{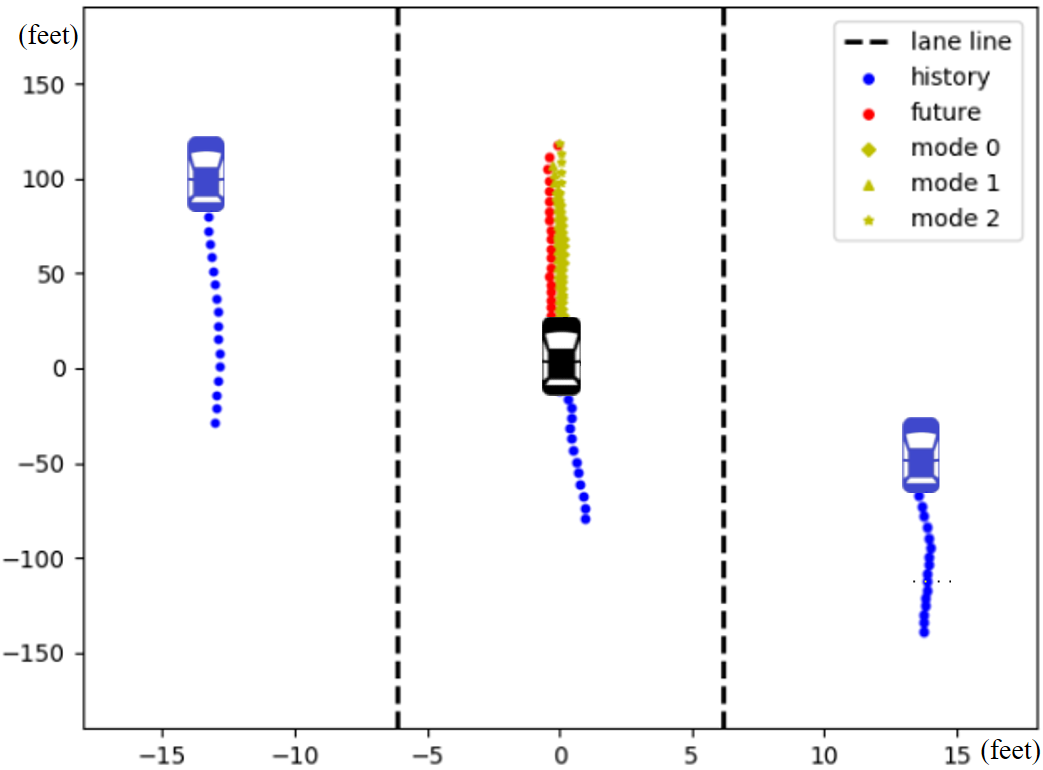}
    }%
    \\
    \subfigure[Under wrong intention.]{
        \includegraphics[width=75mm, trim = 0mm 0mm 0mm 0mm, clip=true]{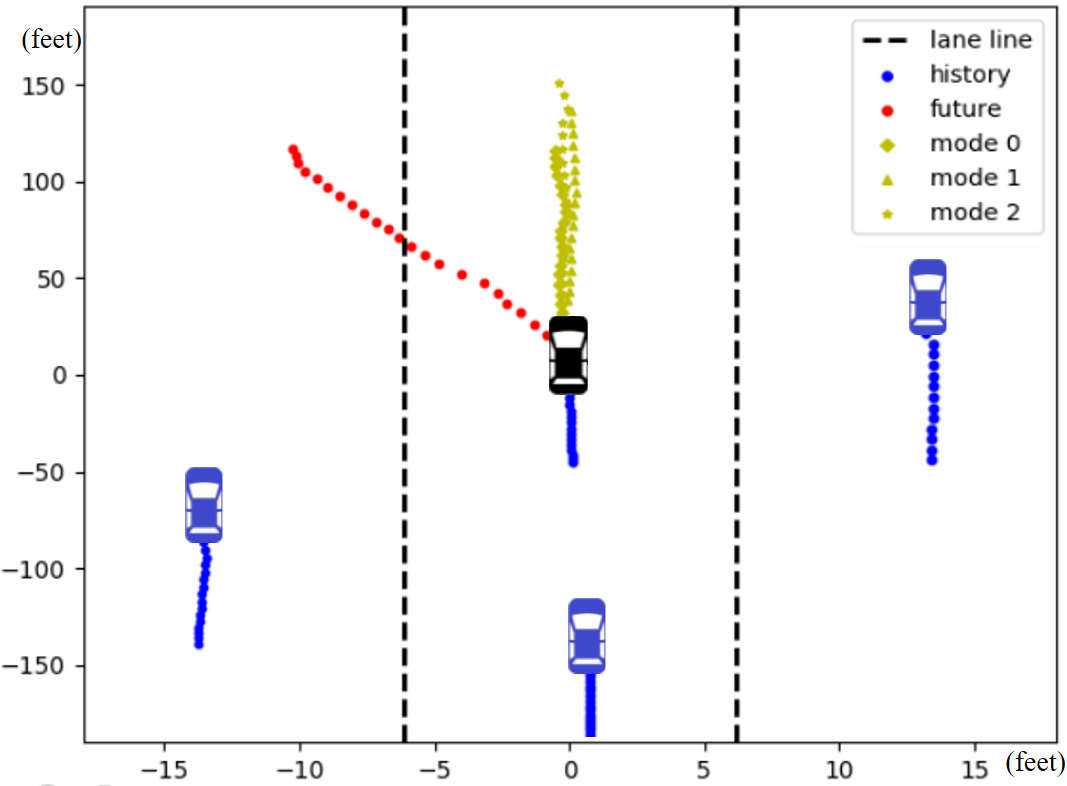}
    }%
    \caption{Prediction with motion modes.}
    \label{Fig6}
\end{figure}

In the next experiment, we only define the motion level for multi-modal trajectories. The number of network outputs is still set to 3. We also adopt the motion mode arbitration process to determine the "winner" during training. 

Qualitative results are shown in Fig.~\ref{Fig6}. We can see that three predicted trajectories are similar, only with minor difference in speed and lateral displacement. It can be considered that they represent different motion modes under a specific intention. If this specific intention is close to the groundtruth, a relatively accurate prediction can be obtained, as shown in Fig.~\ref{Fig6}~(a). Otherwise, the prediction will be poor, as shown in Fig.~\ref{Fig6}~(b). Therefore, without an accurate intention, the motion modes alone cannot yield satisfying predictions.

\begin{figure}[tbp]
    \centering
    \includegraphics[width=75mm]{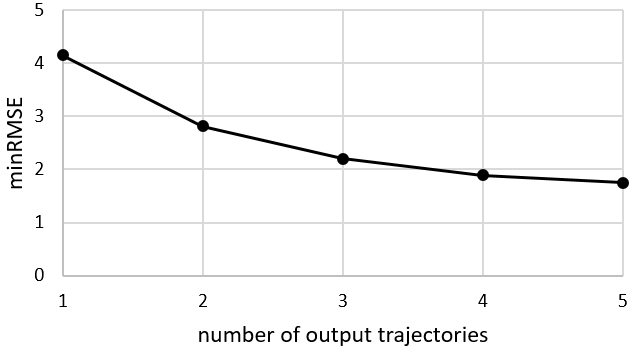}
    \caption{Comparison of prediction performance with different motion mode numbers}
    \label{Fig7}
\end{figure}

Additionally, we evaluate the prediction performance by varied output number of the network. As can be seen from Fig.~\ref{Fig7}, with the increasing number of output trajectories, the accuracy increases as well but the trend slows down. Recall the examples illustrated in Fig.~\ref{Fig6}, we believe that prediction errors are mainly caused by two aspects: wrong intention mode and wrong motion mode. With the increasing number of modes, errors caused by the latter gradually decrease, but the former is still difficult to solve. Hence, the prediction accuracy gradually reaches saturation.

For evaluating the dual-level multi-modal characteristics of vehicle motion, we combine two arbitration processes during training. First, the specific intention is determined through the intention mode arbitration process and then the motion mode of "winner" is determined by the motion mode arbitration process.

\begin{figure}[tbp]
    \centering
    \includegraphics[width=75mm]{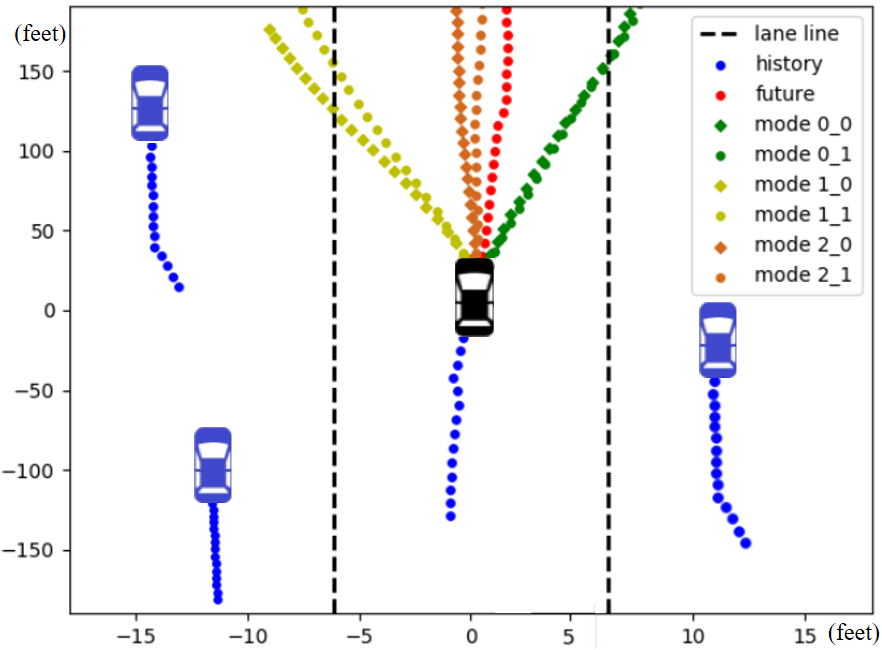}
    \caption{Predictions by combining two arbitration processes.}
    \label{Fig8}
\end{figure}

Qualitative results are shown in Fig.~\ref{Fig8}. As can be seen, the proposed DsMCL training method perfectly combines the advantages of both arbitration processes and the final prediction results can fully consider dual-level multi-modal characteristic of vehicle motion.

\subsection{Quantitative Evaluation}

Here, we compare our CS-LSTM-DsMCL method with the original version of CS-LSTM. Additionally, we adopt two other frameworks for comparison: S-LSTM and S-GAN. The two frameworks utilize the Social Pooling module~\cite{6} and the Pooling module~\cite{8} to substitute the Convolutional Social Pooling module in CS-LSTM respectively. We also adopt our DsMCL training method for both of them and obtain S-LSTM-DsMCL and S-GAN-DsMCL.

\begin{table}[tbp]
\caption{Results of comparison experiments (minRMSE in meters).}
\label{tab:abblation}
	\resizebox{0.98\linewidth}{!}{ %
		\setlength\tabcolsep{5pt}
\begin{tabular}{lllllll}
\multicolumn{2}{l}{Prediction Horizons (seconds)} & 1             & 2             & 3             & 4             & 5             \\
\hline
\multirow{6}{*}{NGSIM}       & CS-LSTM            & 0.51          & 0.99          & 1.46          & 2.04          & 2.89          \\
                             & CS-LSTM-DsMCL       & \textbf{0.51} & \textbf{0.90} & \textbf{1.24} & \textbf{1.67} & \textbf{2.40} \\
                             & S-LSTM             & 0.51          & 0.97          & 1.43          & 1.98          & 2.82          \\
                             & S-LSTM-DsMCL        & \textbf{0.52} & \textbf{0.94} & \textbf{1.29} & \textbf{1.75} & \textbf{2.49} \\
                             & S-GAN              & 0.58          & 1.02          & 1.45          & 1.97          & 2.74          \\
                             & S-GAN-DsMCL         & \textbf{0.51} & \textbf{0.89} & \textbf{1.21} & \textbf{1.62} & \textbf{2.32} \\
\hline
\multirow{2}{*}{HighD}       & CS-LSTM            & 0.08          & 0.18          & 0.30          & 0.44          & 0.61          \\
                             & CS-LSTM-DsMCL       & \textbf{0.08} & \textbf{0.14} & \textbf{0.23} & \textbf{0.35} & \textbf{0.51}\\
                              & S-LSTM             & 0.09          & 0.19          & 0.31          & 0.47          & 0.65          \\
                             & S-LSTM-DsMCL        & \textbf{0.08} & \textbf{0.15} & \textbf{0.25} & \textbf{0.39} & \textbf{0.55} \\
                             & S-GAN              & 0.11          & 0.21          & 0.35          & 0.52          & 0.67          \\
                             & S-GAN-DsMCL         & \textbf{0.10} & \textbf{0.21} & \textbf{0.32} & \textbf{0.44} & \textbf{0.57} \\
\hline
\end{tabular}
}
\end{table}

Results of comparison experiments are reported in Table~\ref{tab:abblation}. It can be found out that the DsMCL training method achieves a significant gain on prediction accuracy for different frameworks on both datasets. This advantage is more evident for longer prediction horizons. It also shows that the DsMCL method is with an excellent generalization ability and not limited to specific network frameworks and dataset.

Moreover, we find out that all methods perform significantly better on HighD dataset than on NGSIM. This may due to the cleaner data, the simpler motion and interaction in the HighD dataset. In experiments, in order to shorten the training time, we only use the first 10 sub-datasets from the huge HighD dataset which recorded 12 times as many vehicles as NGSIM, and these sub-datasets are proven to be sufficient for the training.

In a further experiment, we choose one of the state-of-the-art sampling-based methods for comparison, \ie, the MFP-k~\cite{10}. This compared method learns the multi-modal characteristics of vehicle motion by $k$ latent variables and obtain prediction trajectories by sampling. The MFP-k requires no pre-defined modes and also reports minRMSE scores over 5 samples. Here we compare the performance of MFP-k, the original CS-LSTM and our best method (S-GAN-DsMCL), which respectively represent three different strategies in dealing with multi-modal characteristics of vehicle motion. Test results on the NGSIM dataset by the minRMSE metric are reported in Table~\ref{tab:compare}. From the results, it can be noticed that the S-GAN framework integrated with our proposed DsMCL surpasses the other two methods.
We also find out that the original CS-LSTM with pre-defined multi-modality also performs significantly better than the sampling-based method MFP-k in terms of minRMSE metric (they have not been compared by the same metric before). The reason may be that the predefined modes are good at describing the whole probability space. In contrast, the sampling approach may not be able to cover all possible modes, since modes with high probabilities might be sampled repeatedly while the ones with low probabilities might not be sampled.

\begin{table}[tbp]
\caption{Comparison of three different strategies in dealing with multi-modal characteristics of vehicle motion (minRMSE in meters).}
\label{tab:compare}
	\resizebox{0.98\linewidth}{!}{ %
		\setlength\tabcolsep{5pt}
\begin{tabular}{lllllll}
\multicolumn{2}{l}{{Prediction Horizons (seconds)}}                                         & {1}    & {2}    & {3}    & {4}    & {5}    \\
\hline
% \multirow{6}{*}{NGSIM} & MFP-1                                                                     & 0.54          & 1.16          & 1.90          & 2.78          & 3.83          \\
%                        & MFP-2                                                                     & 0.55          & 1.18          & 1.92          & 2.80          & 3.85          \\
%                        & MFP-3                                                                     & 0.54          & 1.17          & 1.91          & 2.78          & 3.83          \\
\multirow{3}{*}{NGSIM}
                       & MFP-k [10]                                                                   & {0.54} & {1.16} & {1.89} & {2.75} & {3.78} \\
%                        & MFP-5                                                                     & 0.55          & 1.18          & 1.92          & 2.78          & 3.80          \\
                       & CS-LSTM [7]                                                                   & 0.51          & 0.99          & 1.46          & 2.04          & 2.89          \\
                       & \begin{tabular}[c]{@{}l@{}}S-GAN-DsMCL (Ours)\end{tabular} & \textbf{0.51} & \textbf{0.89} & \textbf{1.21} & \textbf{1.62} & \textbf{2.32} \\
\hline
\end{tabular}
}
\end{table}

\section{CONCLUSION}

In this work, we decompose the multi-modal characteristics of vehicle motion into two levels. The first level is the multi-modality of intention from a macro view and the other level is the multi-modality of motion from a micro perspective. Based on that, we propose the Dual-level Stochastic Multiple Choice Learning method. This method does not require human-labeled modes and can make comprehensive probabilistic multi-modal trajectory predictions in a single forward propagation. By experiments on both NGSIM and HighD datasets, our method has proven significant improvement on several trajectory prediction frameworks and achieves state-of-the-art performance.

\section*{ACKNOWLEDGMENT}

This work was supported in part by the National Key Research and Development Program of China under Grant 2018YFB0105103.

%%%%%%%%%%%%%%%%%%%%%%%%%%%%%%%%%%%%%%%%%%%%%%%%%%%%%%%%%%%%%%%%%%%%%%%%%%%%%%%%%

\end{document}